# Non-Invasive Functional-Brain-Imaging with a Novel Magnetoencephalography System

Amir Borna, *Member, IEEE*, Tony R. Carter, Anthony P. Colombo, Yuan-Yu Jau, Jim McKay, Michael Weisend, Samu Taulu, Julia M. Stephen, and Peter D. D. Schwindt

*Abstract*—A non-invasive functional-brain-imaging system based on optically-pumped-magnetometers (OPM) is presented. The OPM-based magnetoencephalography (MEG) system features 20 OPM channels conforming to the subject's scalp. Due to proximity (12 mm) of the OPM channels to the brain, it is anticipated that this MEG system offers an enhanced spatial resolution as it can capture finer spatial features compared to traditional MEG systems employing superconducting quantum interference device (SQUID). We have conducted two MEG experiments on three subjects: somatosensory evoked magnetic field (SEF) and auditory evoked magnetic field (AEF) using our OPM-based MEG system and a commercial SQUID-based MEG system. We have cross validated the robustness of our system by calculating the distance between the location of the equivalent current dipole (ECD) yielded by our OPM-based MEG system and the ECD location calculated by the commercial SQUID-based MEG system. We achieved sub-centimeter accuracy for both SEF and AEF responses in all three subjects.

## I. Introduction

Direct non-invasive brain imaging relies on sensing either the electric field, electroencephalography (EEG), or magnetic field, magnetoencephalography (MEG) [1], or a combination of both [2] outside the skull. These electromagnetic fields are mainly caused by neuronal current sources in the cerebral cortex [1, 3], and finding the precise location, orientation and strength of these neuronal current sources with high spatiotemporal resolution is the holy grail of non-invasive functional brain imaging.

While EEG methods benefit from the simplicity of the instrumentation, they suffer from significant setup time and low spatial resolution (2 cm); this limitation is imposed by the neuronal return currents passing through the skull tissues, which have a low-conductivity profile compared to the surrounding cortex, dura, scalp, and skin tissues. MEG measures signals which are nine orders of magnitude smaller than the earth's magnetic field; hence, it requires sophisticated instrumentations and measurement methods [1]. However, magnetoencephalography has shown spatiotemporal resolution superior to that of the EEG and precision has continued to improve with advancing analysis methods [4, 5], instrumentation [6-11], and systems [12-16].

Traditional instrumentation for MEG data acquisition has been superconducting quantum interference device (SQUID) magnetometers which employ a macroscopic quantum phenomenon in its Josephson junctions [17]. While SQUID systems benefit from mature technology and analysis methods, they face two limiting factors: 1) fixed sensor positions, and 2) high maintenance cost. SQUID sensors operate at ~ 4 K and liquid helium is used to achieve cryogenic temperature. Regular costly maintenance is required to fill the helium reservoir and calibrate the SQUID-based MEG system, although recent advancements in helium recycling [18] is reducing maintenance costs. The main limitation of SQUID-based MEG system, which also stems from the use of cryogens, is fixed sensor position. Due to use of liquid helium a thick Dewar is required to isolate the sensors from the room temperature, hence the fixed position of sensors inside the rigid helmet. In commercial SQUID based MEG systems, the rigid helmet is designed to fit the 95th percentile head size; this fixed helmet size degrades the signal quality for smaller heads, e.g. children [18].

Optically pumped magnetometers (OPMs) offer a new paradigm for MEG measurements [6-11]. While the basic techniques for high sensitivity OPMs were demonstrated in 1969 [19], it took until the early 2000s to realize sub-femtoTesla sensitivity, a sensitivity rivaling that of SQUID [7]. For this OPM, its sensing mechanism typically happens above room temperature and hence does not require cryogenic cooling. Each of the array's sensors can be placed conformal to the individual subject's scalp, maximizing the signal strength.

In this paper, we report on the development of a 20-channel magnetoencephalography system to study complex neural circuits, non-invasively, in human subjects. Optically pumped magnetometers, covered in Section II.A, are used as the magnetic sensors of our MEG system, and the system components are briefly described in Section II.B.. Section III is dedicated to experimental setup and signal processing methods employed to process the collected magnetic fields. To validate the functionality of our MEG system, we have localized somatosensory evoked magnetic fields (SEF) and auditory evoked magnetic fields (AEF) in three adult subjects with sub-centimeter accuracy; the results of neuronal source localization are presented in Section IV. Section V discusses the encountered technical issues; and Section VI concludes the presented work.

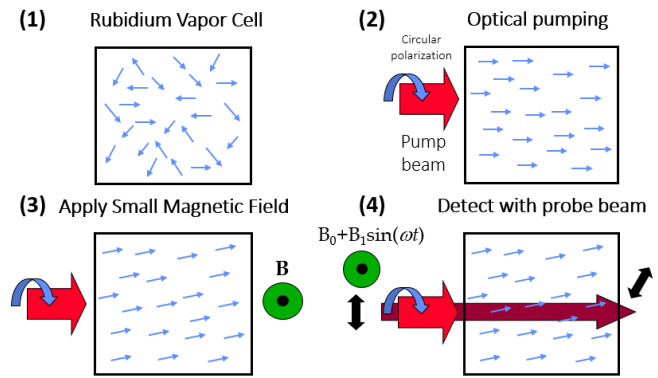

Fig. 1. Optically-pumped-magnetometer's principle of operation: (1) The rubidium atoms of the vapor cell have randomly oriented atomic spins; (2) using the circularly polarized pump laser (795 nm), the spins are aligned in the propagation direction of the pump laser; (3) due to an external magnetic field the atomic spins precess; (4) the precessed atomic spins, through Faraday rotation, change the polarization of the probe beam (780 nm). The polarization change of the probe beam is proportional to the sensed magnetic flux density.

## II. MATERIAL

### A. Magnetic Sensor

The principle of operation of low-field optically pumped magnetometry is illustrated in Fig. 1. These sensors operate in the spin-exchange-relaxation-free (SERF) regime [7]. At the heart of these optically pumped magnetometers is a vapor cell where the light-atom interaction takes place. The vapor cells in our system contain a small droplet of rubidium ($^{87}$Rb) and by heating up the vapor cells to a temperature of ~ 180 °C, we achieve a high density of rubidium atoms (~$10^{13}$ cm$^{-3}$). A circularly polarized pump laser, tuned to rubidium's D1 spectral resonance line (795 nm), aligns the atomic spins, and puts the rubidium atoms in a magnetically sensitive state. The rubidium atoms precess in the presence of an external magnetic field, and a change in angle of the atomic spins changes the polarization angle of a linearly polarized laser beam, i.e. the probe beam (780 nm). The probe-beam is coaligned with the pump-beam, and as it passes through the rubidium vapor, its polarization rotates through Faraday rotation [20]. The change in orientation of the probe beam's polarization is proportional to the sensed magnetic flux density [8]. The magnetic sensor of our system is the OPM, described in [8] (Fig. 2). Our sensor features 4-channels formed at the intersection of four laser beams with the vapor cell's rubidium atoms. The four laser beams have a separation of 18 mm with a full width at half maximum (FWHM) of 2.5 mm. Based on the height of vapor cell, 4 mm, each channel has a sensing volume of $4\ mm \times \pi \times (\frac{FWHM}{2})^2 = 20\ mm^3$. To achieve thermal isolation between the vapor cell walls, operating at ~ 180 °C, and the subject's scalp we have added an extra layer of polyimide insulation with a thickness of 3 mm. Hence, the total distance between the subject's scalp, i.e. source space, and the geometric center of the channels sensing volume is 12 mm. Using on-sensor coils, a 1 kHz modulating magnetic field is applied to each sensor's vapor cell in either vertical or horizontal direction, referenced to the sensor's frame of reference. The modulating magnetic field defines the sensing axis for the sensors, and hence renders a vectoral measurement.

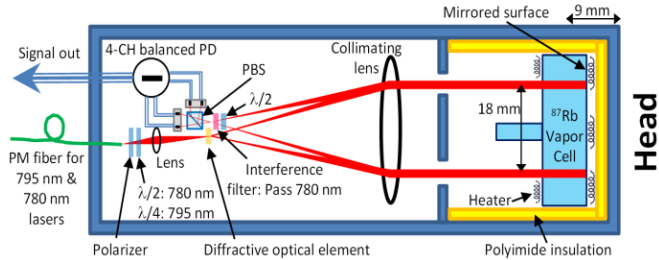

Fig. 2. The OPM sensor's schematic [8]. PBS: polarizing beam splitter; PM: polarization maintaining, PD: photodiode, λ/2: half wave plate, λ/4: quarter wave plate.

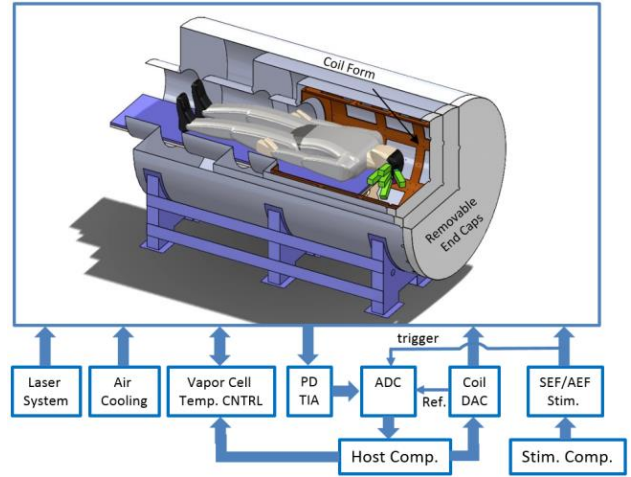

Fig. 3. The MEG system block diagram [11]. PD TIA: the transimpedance amplifier which amplifies the currents from the sensors' photo diodes; temp. CNTRL: temperature control; ADC: analog-to-digital converter; DAC: digital-to-analog converter; SEF/AEF Stim.: somatosensory/auditory stimulation; Ref.: 1 kHz reference for the software lock-in amplifier.

### B. System Components

In [14], we discuss in detail design and characterization of our OPM MEG system. The system block diagram is shown in Fig. 3. Multiple subsystems were developed: 1) a person-sized magnetic shield with magnetic field control; the shield has a length of 269 cm and an external diameter of 140 cm. There are 18 coils embedded in the shield to null the remnant DC magnetic field around the sensor array; 2) a laser and optical system to provide light to five sensors; 3) a data acquisition (DAQ) system to digitize the sensors; 4) custom-electronics for closed-loop vapor cell temperature control; 5) custom-electronics to operate the sensors and null the remnant DC magnetic field inside the shield; and 6) customized LabVIEW (National Instruments, US) software for data acquisition and system control.

### C. Signal Path

The signal path's block diagram is depicted in Fig. 4. The electric-current from the sensor's photodiodes is sent to a transimpedance amplifier (TIA) (QuSpin, Inc.) and the resulting output voltage is digitized by the MEG system's data acquisition module at 100 kS/s with 24-bit resolution. The digitized raw signal is modulated at 1 kHz, and its AC amplitude contains the light polarization signal which is proportional to magnetic flux density sensed by the OPM. A software lock-in amplifier (LIA) implemented in LabVIEW (National Instruments, US) demodulates the sensor signal and stores the magnetic flux density to hard drive at a decimated rate of 1 kS/s.

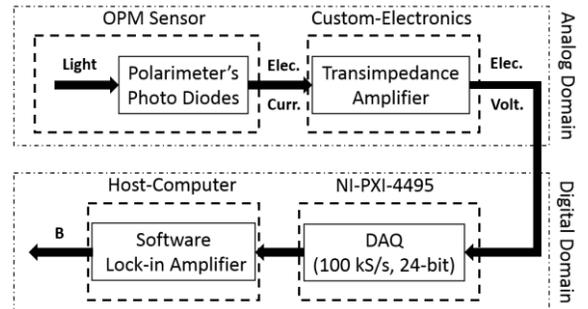

Fig. 4. Signal path: the probe beam's polarization is converted into electrical current by the sensor's polarimeter; the amplified electrical current is digitized by a sampling rate of 100 kS/s and a resolution of 24-bit using commercial data acquisition cards; using custom-designed software lock-in amplifier (LIA) the sensed magnetic flux density is calculated and stored on the host computer's hard drive.

## III. METHODS

### A. Experimental Setup

For conducting OPM-MEG measurements, the subject lies in the supine position inside the custom-designed magnetic shield. During the OPM-MEG experiment, the 20-channel OPM array covers the left hemisphere of the subject's head. Depending on the type of MEG experiment, i.e. AEF vs. SEF, the subject is guided to tilt his/her head such that the OPM array covers either the auditory or somatosensory cortices.

We conducted the MEG experiments on three healthy male subjects aged between 38 and 43 years old. We defined the accuracy of our MEG system as the distance between the locations of the localized neuronal current sources yielded by our OPM-based MEG system and a commercial SQUID-based MEG system. For comparison purposes, the MEG experiments were conducted with the exact same protocols using a 306-channel Elekta-Neuromag SQUID system (Elekta, Sweden) located in a magnetically shielded room (MSR) at the Mind Research Network (Albuquerque, NM). The protocols of the MEG experiments were approved by the Human Studies Board (HSB) of Sandia National Laboratories and Chesapeake Institutional Review Board (IRB).

*1) Auditory Evoked Magnetic Fields (AEF)*

To stimulate the auditory cortex of the subjects, they were presented with a series of standard 1 kHz tones and rare 1.2 kHz tones. The pulse duration of both tones was set at 100 ms and they were presented with random intervals around 1.1 s. Non-magnetic Insert-Earphones (Etymotic Research, Inc., US) deliver the audio stimuli to the subject's ears. The earphones receive the audio signal from the host computer's audio card which is controlled by the stimulus delivery program Presentation (Neurobehavioral Systems, US). Apart from the audio card, the stimulus delivery program also controls the computer's parallel port (LPT). The LPT port is used to send two trigger signals, associated with standard/rare tones, to the MEG system's data acquisition module. Synchronization between the presented audio signals and the trigger signals is essential as timing jitter leads to data corruption. We have a maximum measured jitter of 200 µs. The subject is presented with a total of 360/150 audio pulses for the standard/rare tones.

*2) Somatosensory Evoked Magnetic Fields (SEF)*

To stimulate the somatosensory cortex of the subjects, we sent current pulses to the subject's median nerve on the right wrist through two 8-mm felt pads, spaced 25 mm apart. The unipolar stimulus signal has a pulse-width of 200 µs and its amplitude was set, individually for each subject, according to the 2-cm thumb twitch response. The stimulus delivery program Presentation (Neurobehavioral Systems, US), running on the host computer, controls the timing of current stimulus by sending a trigger signal through the computer's parallel port (LPT) to the stimulator module. The stimulator module is a commercial constant-current high-voltage peripheral stimulator, DS7A (Digitimer, United Kingdom). The stimulator's trigger is also routed to the MEG system's data acquisition module which collects the MEG data simultaneously with the trigger signal. The stimulus delivery program sends 400 trigger pulses, with random intervals around 1 s, to the DS7A stimulator.

### B. Signal Processing Pipeline

The signal processing pipeline processes the stored magnetoencephalography data and constructs the forward model. For our MEG signal processing tasks, we use Fieldtrip [21], a Matlab (MathWorks, Inc.) toolbox offering advanced analysis methods for MEG.

*1) Processing the Magnetoencephalography Data*

The digitized, time-domain magnetoencephalography data is bandpass filtered from 0.5-150 Hz. The power of the filtered signal is used as a criterion to detect corrupted channels, as in our system corrupted channels have large background noise; After removing the malfunctioning channels, the segments of the continuous, bandpass-filtered signal, which are contaminated with *1)* discontinuities, *2)* muscle/movement artifacts, and *3)* electrooculogram (EOG) artifacts, are identified and subsequently removed from the continuously recorded data. Using the trigger signals recorded alongside the MEG data (Section II.1.A/B) the trials are defined for the clean data segments. Depending on the subject movements during the experiment, this step might remove as many as half the trials. The remaining trials were averaged relative to the onset of trigger. To further clean the MEG data, we ran independent component analysis (ICA) on a continuous segment of data with no large noise component such as muscle or movement artifacts; to have a robust ICA decomposition we choose this segment to be at least 30 s long [22]. Other than the unmixing matrix, the ICA also identified large noise components (virtual channels) such as Magnetocardiogram (MCG), 60 Hz, and shield vibration artifacts. The unmixing matrix yielded by the ICA procedure was used to remove the noisy ICA components from the time-locked sensor-level channels.

*2) Constructing the Forward Model*

The forward model in the context of functional brain imaging calculates the sensor-level magnetic topography, emanating from a specific neuronal current source [3]. The first step in calculating the forward model is finding the relative position of brain anatomy and the sensors, a procedure commonly referred to as coregistration of MRI and MEG data. The basic assumption in MEG/MRI coregistration is that the subject's brain does not shrink or move inside the skull between the MRI and MEG sessions.

Four head position indicator (HPI) coils were fixed on each subject's scalp prior to running the AEF or SEF experiments.

The positions of the HPI coils along with the subject's head shape were digitized using a Fastrack digitizer (Polhemus, US). Before presenting the subject with stimuli, a sequence of AC currents with a peak-to-peak amplitude of 2 mA and a frequency of 5.1, 5.2, 5.3, and 5.4 Hz are applied sequentially to the HPI coils. Even though the coils are activated individually the uniqueness of their frequencies helps identify their location based on the frequency of the recorded magnetic field. The magnetic field of these coils and their activating signal are collected with a sampling frequency of 100 kS/s; however, they are stored on the hard drive with a decimated sampling rate of 1 kS/s. The recoded channels are time-locked (averaged) to their activating signal; the peak amplitude of the resulting time-locked channels are used for dipole fitting [23]. Using the Fieldtrip's dipole fitting routine, the HPI coils positions are determined. The coordinates of the HPI coils are used for coregistration of MEG and MRI data. We use single-shell forward model [24] provided by the Fieldtrip package. The single-shell forward model yields an accuracy comparable to that of the boundary element methods (BEM) for MEG signals but avoids the tedious computations of BEM forward model.

### C. Neuronal Current Source Localization Methods

The inverse problem in the context of functional brain imaging refers to finding the locations and orientations of neuronal current sources underlying a measured sensor-level magnetic spatial topography [3]; this is an inherently ill-posed problem with more unknown parameters than known ones [1]. The precision of neuronal current source localization method depends on the quality of the sensor-level MEG data channels, stability of the magnetic sensors, and the robustness of the forward model. We apply two different methods to localize neuronal current sources activated in the AEF and SEF experiments: equivalent current dipole (ECD) fitting and linearly constrained minimum variance (LCMV).

In both these methods, the (neuronal current) source is modeled as an equivalent current dipole; the neurophysiological reasoning behind this assumption is that activated pyramidal cells in the cortical regions are aligned side by side and the generated magnetic field is measured at a distance by the sensors outside the skull [3].

#### 1) Equivalent Current Dipole Fitting

We used Fieldtrip [21] to implement our dipole fitting routine. We defined the grid points on the whole brain region, rather than cortical regions. With a resolution of 0.5 cm, we complement our grid scanning procedure with non-linear search to fine-tune the position and orientation of an optimal single current dipole. For AEF, even though there are two simultaneously activated bilateral cortical regions, it is safe to assume that a single current dipole is an adequate source model; this assumption is based on the fact that our OPM sensor array is located on the left hemisphere and, due to large distance, does not sense the magnetic field stemming from the activated auditory cortex in the right hemisphere. For SEF, we stimulate the median nerve of the right wrist, hence the activated somatosensory cortex is in the left hemisphere only where the OPM sensor array is located.

#### 2) Adaptive Distributed Source Imaging

We used Fieldtrip [21] to implement linearly-constrained-minimum-variance (LCMV) [25]. The grid resolution was set at 1 cm. LCMV designs an optimal spatial filter such that it passes the response of the filter's focus point while minimizing the variance at the filter output. Using the methods of Lagrange multipliers, LCMV yields the variance (strength) of the current dipole located at the grid points. To alleviate the effect of noise bias toward deep brain structures (furthest from the sensor array), we normalized the dipole strength using noise-covariance. It is common to calculate the noise-covariance matrix using the empty room (no subject) measurement; however, we used the pre-stimulus intervals to calculate the noise covariance matrix. In LCMV highly correlated distant sources cancel each other; in AEF experiments there are two highly correlated bilateral cortical regions which can potentially prohibit the application of LCMV. However, for reasons stated in Section III.C.1, we can safely employ LCMV to calculate the neuronal activities for our AEF experiment.

## IV. RESULTS

Our OPM sensors, operating in the person-sized magnetic shield, show a sensitivity of 10-15 fT/rt-Hz in the magnetometry mode and an inherent noise level of 3-5 fT/rt-Hz in the gradiometry mode; the bandwidth of these sensors is 85-95 Hz [14] which is sufficient to capture the temporal features of auditory and somatosensory evoked magnetic fields [1].

### A. Forward Model Construction

Fig. 5-a shows the time-domain waveform of a single channel as it senses the magnetic fields emanating from four HPI coils activated sequentially. After averaging the time-locked data for all the channels, a single cycle waveform, Fig. 5-b, is generated. The linear range of our sensors is ~ 1.5 nT [8]; the magnetic field from HPI coils can exceed the upper sensing range and subsequently cause distortion in the channels located immediately adjacent to the coils. As an example, in Fig. 5-b, the two channels with largest amplitude suffer from distortion. It is essential to remove these channels before localizing the position of HPI coils, as distortion is not accounted for in the forward model. We feed the time-locked channels to the dipole fitting routine [23] to localize the location of the four HPI coils. Using the calculated position of HPI coils and their digitized counterparts, we coregister the MRI and the MEG data.

To achieve accurate localization, the sensors should cover the spatial patterns generated by the targeted neuronal sources. Due to limited number of sensors, our OPM array is not a full-head MEG system. Hence, the array could cover either the somatosensory or auditory cortex. Before each experiment the positions of the HPI coils were adjusted to cover the cortex of interest (auditory vs. somatosensory) and the subject was asked to tilt his/her head such that the targeted neuronal cortex was covered by the array. Fig. 6 shows the location of the sensors for SEF and AEF experiments. The brain tissue, extracted by segmenting the MRI, is used to generate a single-shell forward

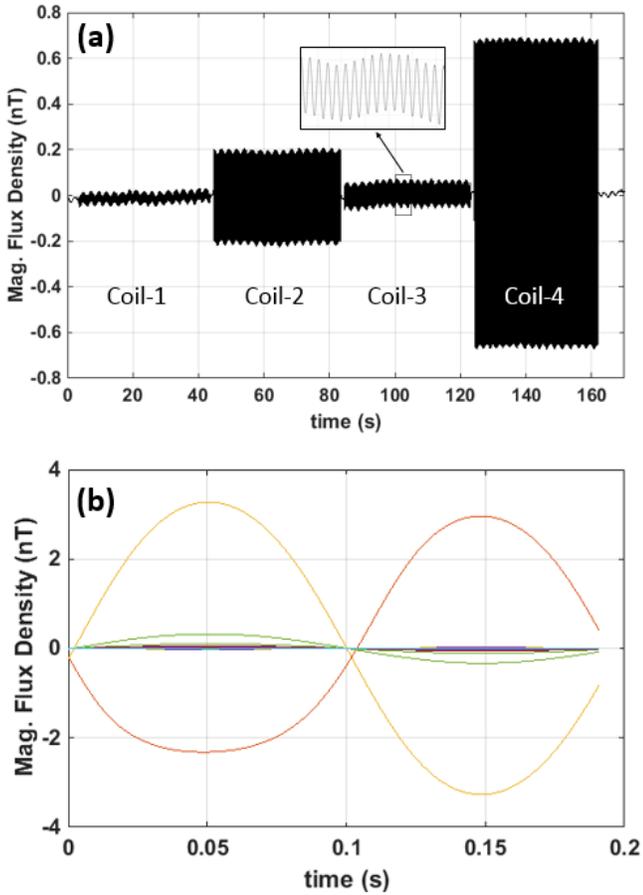

Fig. 5. Time-domain waveforms of head position indicator (HPI) coils: (a) the raw waveform of a single channel for all the four individually activated HPI coils, and (b) time-locked (average) waveform of all the channels for coil-1.

model [24].

### B. MEG Signal Processing

After bandpass filtering the continuously recorded MEG data, the continuous data is segmented into trials (epochs) using the trigger signal (Section II.1.A/B). Trials corrupted by movement, muscle, or EOG artifacts are discarded. However, there are still artifacts in data which were not captured by artifact rejection methods. Independent component analysis (ICA) can identify these noise components. Fig. 7 shows the ICA components of an SEF response; the first six components are unwanted interference and do not contribute to the SEF response. After eliminating these components, the remaining trials were reconstructed and averaged. Fig. 8 shows the effect of running ICA on the SEF response; the raw data has a large component at around 100 ms which dominates the SEF's M20 response. Using ICA, this component is successfully removed from the SEF response and in Fig. 8-b M20 response is clearly visible. Fig. 9 shows the ICA-cleaned, time-domain, time-locked MEG data for SEF (a-c) and AEF (d-f) experiments on all three subjects.

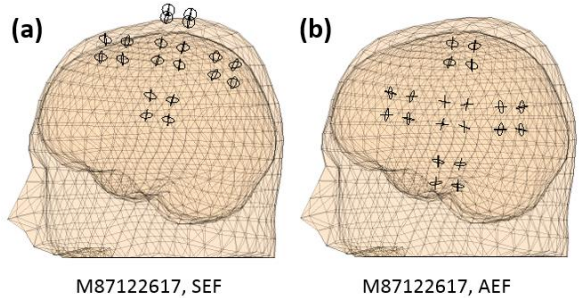

Fig. 6. Coregistration of MRI and MEG data for subject M87122617: (a) coregistration for the SEF experiment, and (b) coregistration for the AEF experiment.

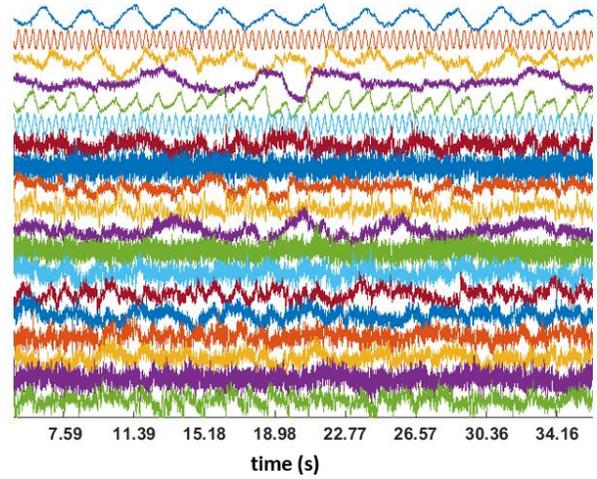

Fig. 7. ICA components of the somatosensory evoked magnetic fields in subject M87172872; the first six components are noise and should be eliminated.

### C. Measured Scalp Magnetic Topography

The ICA-cleaned, time-domain, time-locked MEG data (Section IV.C) was used to create spatial topographies for AEF and SEF responses. It is common to create these field-maps at M20 peak (~ 20 ms) for SEF and M100 peak (~ 100 ms) for AEF data. Fig. 10 shows the spatial topographies for SEF and AEF responses of all subjects; the small circles show the location of the OPM channels. The variation between subjects' field-maps is due to different positions of the subjects' heads with respect to the sensor array.

### D. Linearly Constrained Minimum Variance (LCMV)

To localize the neuronal response, we use LCMV implemented in Fieldtrip [21]. The covariance of the sensor outputs is calculated using a continuous segment containing pre-stimulus and post-stimulus intervals. To normalize the neural activity, we calculate the noise covariance using pre-stimulus interval (-100 ms to 0). Fig. 11 shows the normalized neural activity index of subject-3 for AEF (a-b) and SEF (c-d);

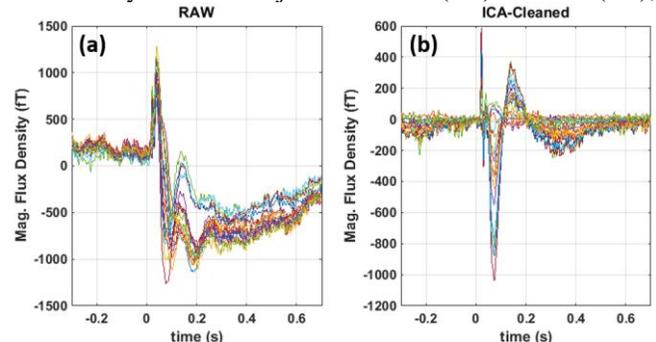

Fig. 8. Time-locked somatosensory evoked magnetic fields in subject M87172872: (a) time-locked raw data before ICA, and (b) time-locked ICA-cleaned data.

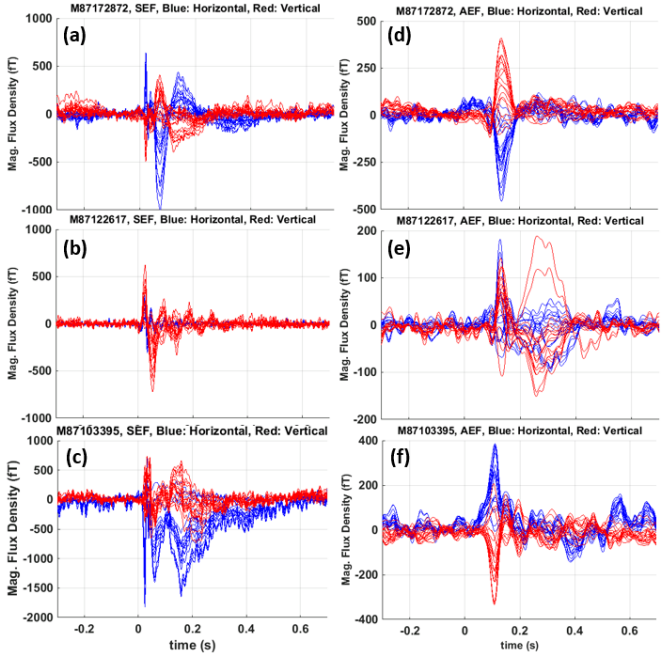

Fig. 9. Time-domain, time-locked data for SEF (a-c) and AEF (d-f) for all three subjects.

the auditory and somatosensory cortices show the highest activities for AEF and SEF experiments, respectively.

*E. Equivalent Current Dipole (ECD)*

For each subject the SEF and AEF experiments are replicated and measured using a commercial Elekta-Neuromag SQUID-based MEG system. On the SQUID-based MEG system, the neuronal responses of SEF and AEF experiments are localized using the dipole-fitting method. To evaluate the robustness of our system, we implemented dipole-fitting in our signal processing pipeline using Fieldtrip [21]. Using ICA-cleaned, time-domain, time-locked SEF/AEF waveforms (Fig. 9) we localized the neuronal responses for SEF/AEF experiments at M20/M100 peaks. We used 2 ms around the response peak to average the OPM channel signal. Fig. 12 shows the dipole locations calculated using dipole-fitting [23] on the SEF data of all subjects collected with our OPM-based MEG system; Fig. 13 shows the dipole locations for the AEF response of all subjects. Table I compares the calculated location of the equivalent current dipole using our OPM-based MEG system to the location provided by the SQUID-based MEG system; for both SEF and AEF responses of all subjects, the error is sub-centimeter.

## V. ANALYSIS AND DISCUSSION

During the SEF experiments using our OPM-based MEG system we encountered a large peak at 100 ms, Fig. 14-a, with a frequency content of less than 20 Hz. The large component at 100 ms was observed in all the subject's SEF response and could reach an amplitude of 20 pT. However, the amplitude of this artifact varies significantly inter-subject and inter-session. Due to large amplitude of this signal, the SEF's M20 and M30 responses were overwhelmed. Bandpass filtering does not efficiently remove this artifact as its large amplitude induces ringing. We isolate this artifact from the SEF response using

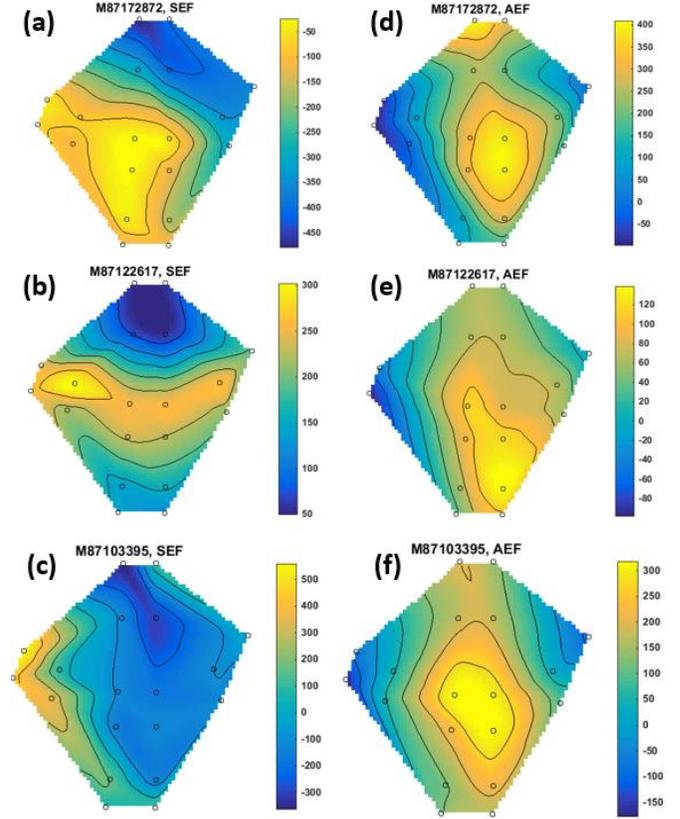

Fig. 10. Spatial topographies for SEF (a-c) and AEF (d-f) for all three subjects. M20 and M100 peaks are used to create the field-maps for SEF and AEF, respectively. The field-maps show the measured magnetic flux density (fT).

independent component analysis (ICA) as shown in Fig. 14-b. The large-amplitude signal is not measured by the SQUID-based MEG system, hence we speculated that its origin is not

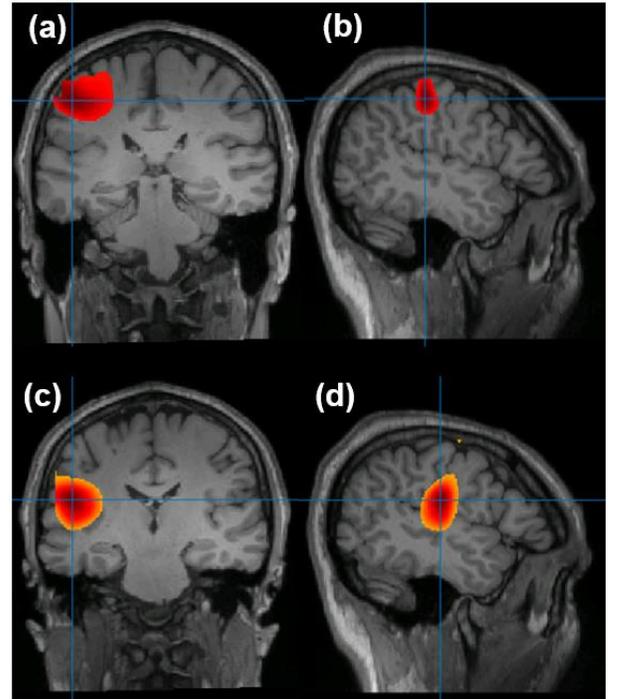

Fig. 11. Localization using LCMV on M87103395 MEG data: (a-b) Coronal and Sagittal planes of the SEF data, (c-d) Coronal and Sagittal planes of the AEF data.

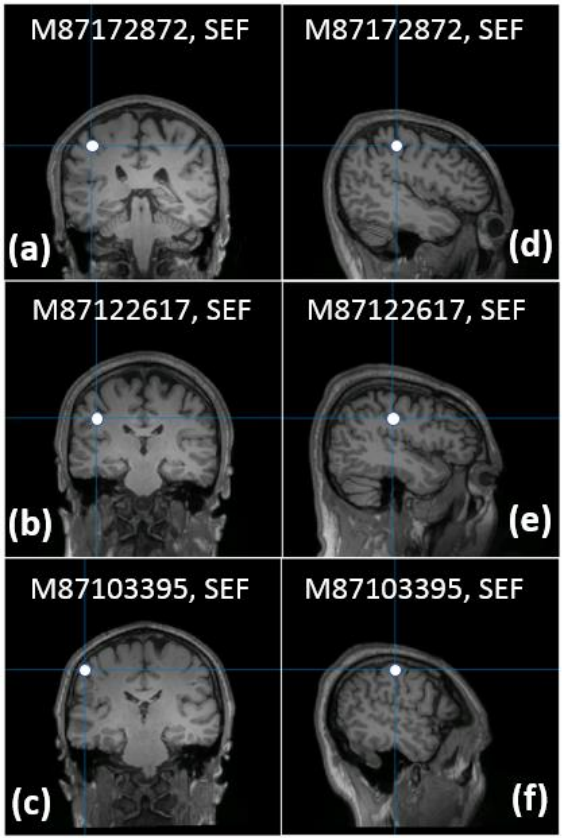

Fig. 12. Localization of the SEF's M20 response using equivalent current dipole fitting: (a-c) Coronal and (d-f) Sagittal planes of the dipole location. The white dot shows the calculated dipole location.

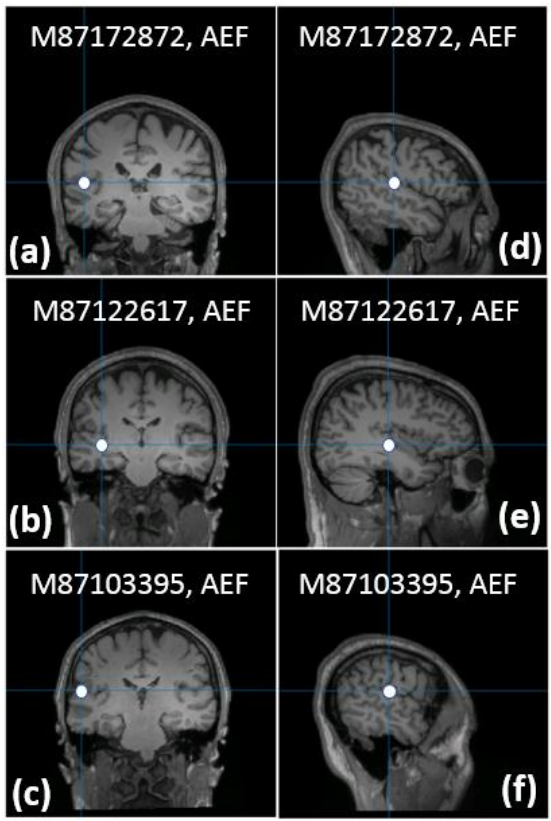

Fig. 13. Localization of the AEF's M100 response using equivalent current dipole fitting: (a-c) Coronal and (d-f) Sagittal planes of the dipole location. The white dot shows the calculated dipole location.

TABLE I
COMPARISON OF THE AEF/SEF SOURCE LOCALIZATION ACCURACY IN THREE DIFFERENT MALE ADULT SUBJECTS

| Subject ID | AEF Err. (mm) | SEF Err. (mm) |
|---|---|---|
| M87172872 | 9.5 | 3.9 |
| M87122617 | 7.1 | 7.5 |
| M87103395 | 6.2 | 9.7 |

Ground truth is the source location provided by the commercial Elekta-Neuromag SQUID-based MEG system.

from the subject's head. The artifact's onset is before the M20 response; hence it has a different propagation path than the subject's median nerve. We speculate that the small movement from the SEF's thumb twitch induce movement in magnetic shield relative to the OPM array which is responsible for the up-to-20-pT-amplitude artifacts at 100 ms. In future we plan to solve this issue by substantially stiffening the flat endcaps of the shield and by adding dampening materials between the shield layers.

The neuronal localization algorithms employed here, i.e. dipole-fitting and spatial filtering, assume characteristics of the array channels, e.g. gain and sense-angle (the vector component of the magnetic field measured by the OPM channel), are constant throughout all the trials. If channel position, gain, or sense-angle varies between trials, it can induce significant localization error. We calibrate our OPM array before each experiment by measuring the channels gain and sense-angle. The gain of OPM channels is mainly proportional to the power of the lasers in the sensor and the Rb vapor density, which can drift. Laser power drift can be minimized by a more stable mechanical design that limits drift in coupling to optical fibers or by adding active feedback. Ideally the channels' sense-angle is dependent only on the orientation of on-sensor coils and their currents; hence, constant throughout the experiment. However, we observed a significant drift in the sense-angle (~ 2.5 °) over the course of a couple of hours (Fig. 15). This drift was attributed to drifts in the pump laser causing a varying light shift, which produces a fictitious magnetic field along the direction of the pump laser. Originally, there were two pump-lasers separated by 20 GHz and centered around the rubidium's D1 transition; due to different polarization rotations induced by the polarization maintaining optical fibers, the two lasers were experiencing a different rotation in the fiber and then a different amplitude after passing through the sensor's polarizer. The variation between the two pump-lasers' power level in the sensor created the drift in sense-angle. By switching from a double-pump-laser scheme to a single-pump-laser scheme we were able to reduce the sense-angle drift by an order of magnitude as shown in Fig. 15. Only then, were we able to localize the neuronal current sources with sub-centimeter accuracy. In the future with improvements to the optical system, we plan to switch back to the two-pump laser scheme since it offers a factor of two larger gain in the OPMs.

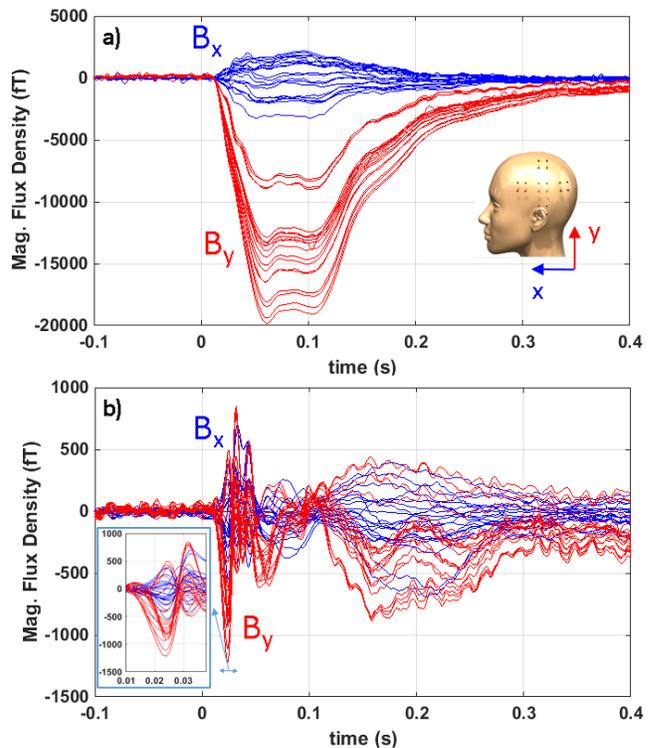

Fig. 14. Evoked response due to median nerve stimulation showing both the horizontal and vertical field components from all 20 channels. a) the raw data including the shield artifact, and b) the SEF data cleaned by ICA where the large 100 ms component is removed from the data. The inset in (b) depicts the M20 and M30 components. Filter: bandpass 0.1 Hz to 150 Hz.

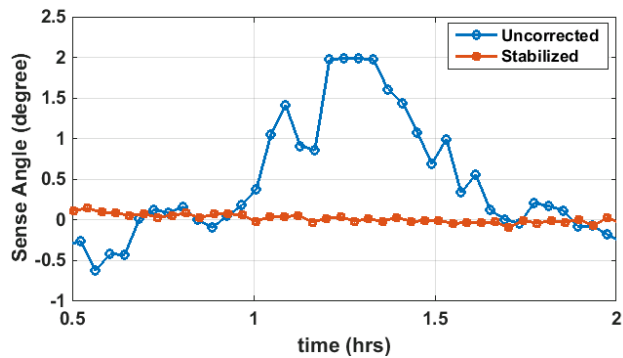

Fig. 15. The sense-angle measured over the course of two hours using uncorrected and corrected laser systems. The single frequency pump laser has reduced the sense angle variation by more than an order of magnitude.

## VI. Conclusion

We have developed a 20-channel OPM-based MEG system using our custom-designed OPM sensor. Due to proximity (12 mm) of the magnetic sensor channels to the source space, i.e. subject's scalp, this MEG system offers an enhanced spatial resolution as it can capture finer spatial features compared to traditional SQUID-based MEG systems. Using our OPM-based MEG system we have conducted auditory evoked magnetic field (AEF) and somatosensory evoked magnetic field (SEF) experiments on three subjects. Using a commercial Elekta-Neuromag SQUID-based MEG system as a reference, the OPM-based MEG system yielded neuronal current source localization results with sub-centimeter accuracy for both responses in all three subjects.

Acknowledgment

The authors thank Dr. John Mosher for helpful discussion on operation of the OPM array and conducting MEG experiments; they also thank Kim Paulson for help in collecting the SQUID MEG data, Jeff Hunker for building the 18-coil system for the OPM MEG system, and QuSpin Inc. for building the multichannel transimpedance amplifier. Sandia National Laboratories is a multimission laboratory managed and operated by National Technology & Engineering Solutions of Sandia, LLC, a wholly owned subsidiary of Honeywell International Inc., for the U.S. Department of Energy National Nuclear Security Administration under contract DE-NA0003525. This paper describes objective technical results and analysis. Any subjective views or opinions that might be expressed in the paper do not necessarily represent the views of the U.S. Department of Energy or the United States Government. Research reported in this publication was supported by the National Institute of Biomedical Imaging and Bioengineering of the National Institutes of Health under award numbers R01EB013302 and R56EB013302. The content is solely the responsibility of the authors and does not necessarily represent the official views of the National Institutes of Health.